\documentclass[11pt,twoside]{article}
\usepackage{asp2010}

\resetcounters

\begin{document}

\title{The Incidence of Magnetic Fields in Massive Stars: \\ An Overview of the MiMeS Survey Component}
\author{J.H. Grunhut$^{1,2}$, G.A. Wade$^2$, and the MiMeS Collaboration
\affil{$^1$Dept. of Physics, Queen's University, Kingston, Ontario, Canada}
\affil{$^2$Dept. of Physics, Royal Military College of Canada, Kingston, Ontario, Canada}}

\begin{abstract}
With only a handful of known magnetic massive stars, there is a troubling deficit in the scope of our knowledge of the influence of magnetic fields on stellar evolution, and almost no empirical basis for understanding how fields modify mass loss and rotation in massive stars. Most remarkably, there is still no solid consensus regarding the origin physics of these fields - whether they are fossil remnants, or produced by contemporaneous dynamos, or some combination of these mechanisms. This article will present an overview of the Survey Component of the MiMeS Large Programs, the primary goal of which is to search for Zeeman signatures in the circular polarimetry of massive stars (stars with spectral types B3 and hotter) that were previously unknown to host any magnetic field. To date, the MiMeS collaboration has collected more than 550 high-resolution spectropolarimetric observations with ESPaDOnS and Narval of nearly 170 different stars, from which we have detected 14 new magnetic stars.
\end{abstract}

\section{Introduction}
Massive stars are those stars with initial masses above about 8 times that of the sun, eventually leading to catastrophic explosions in the form of supernovae. These represent the most massive and luminous stellar component of the Universe, and are the crucibles in which the lion's share of the chemical elements are forged. These rapidly-evolving stars drive the chemistry, structure and evolution of galaxies, dominating the ecology of the Universe - not only as supernovae, but also during their entire lifetimes - with far-reaching consequences. 

Although the existence of magnetic fields in massive stars is no longer in question, our knowledge of the basic statistical properties of massive star magnetic fields is seriously incomplete. 

Prior to the MiMeS Project only a handful of magnetic early B-type stars were known, the majority of these being the lower-mass helium-strong stars \citep[e.g.][]{boh87}, along with only 3 proposed magnetic O-type stars \citep[e.g.][]{don06, bour08}. The majority of these stars host magnetic fields organized on global scales, with a strong dipole component.

Due to the difficulty in measuring magnetic fields in massive stars (e.g. due to the broad and relatively few optical spectral lines) the relatively weak magnetic fields in these stars have remained undetected by previous generations of instrumentation. Many new magnetic massive stars have been reported in the recent literature based on low-resolution spectropolarimetric observations obtained primarily with the VLT's FORS instruments. A significant number of those claims have not been confirmed by independent observations, and even re-examination of the published data \citep[e.g.][]{silv09, bagnulo11, shultz11}. In light of these issues it is challenging to interpret the relatively few existing surveys of magnetism in massive stars. This represents a fundamental gap in our understanding of stellar magnetism and the physics of hot stars.

The Magnetism in Massive Stars (MiMeS) Project represents a comprehensive, multidisciplinary strategy by an international team of recognized researchers to address the ``big questions" related to the complex and puzzling magnetism of massive stars. In 2008, the MiMeS Project was awarded ``Large Program" (LP) status by both Canada and France (PI G. Wade) at the Canada-France-Hawaii Telescope (CFHT), where the Project was allocated 640 hours of dedicated time with the ESPaDOnS spectropolarimeter from late 2008 through 2012. Since then the MiMeS consortium was awarded additional LP status with the Narval spectropolarimeter at the Bernard Lyot Telescope in France (a total of $\sim$500 hours, PI C. Neiner) and the HarpsPol at ESO's 3.6\,m telescope at La Silla, Chile (a total of $\sim$300 hours, PI E. Alecian). In addition to these LPs, the MiMeS Project is supported by numerous PI programs from such observatories as the Anglo-Australia Telescope, Chandra, Dominion Astrophysical Observatory, HST, MOST, SMARTS, and the Very Large Telescope (VLT).

The structure of the ESPaDOnS MiMeS LP includes the ``Survey Component" (SC), which includes over 385 hours dedicated to searching for magnetic fields in $\sim250$ stars with no prior detection of a magnetic field, with the ultimate goal to provide critical missing information about field incidence and statistical field properties from a large sample of massive stars. The majority of the SC targets were chosen such that they have $V<9$ and a projected rotational velocity $v\sin i<300$\,km\,s$^{-1}$ to ensure sensitivity to relatively weak large-scale fields.

\section{Observations}
The primary instruments used for the SC are the bench-mounted, high throughput, fibre-fed, high-resolution ($R\sim65000$) ESPaDOnS spectropolarimeter and its twin Narval. The spectrograph is both thermally and vibrationally isolated to reduce systematic effects, which could result in spurious magnetic signatures. The spectrograph is capable of obtaining near complete coverage of the optical spectrum (370 to 1050\,nm) in a single exposure in all four Stokes parameters. Our exposure times are chosen to statistically provide signal-to-noise ratios (S/N) high enough to detect dipole magnetic fields with a range of surface polar field intensities from 100\,G to 2\,kG depending on the stellar, physical and spectral properties.

To diagnose the magnetic field, the SC relies on the magnetic splitting of photospheric spectral lines (the Zeeman effect). Since the magnetic splitting observed in unpolarized light is difficult to detect in hot stars due to their intrinsically broad spectral lines (from thermal, turbulent or rotational broadening), we rely on the Zeeman signatures that are produced in circularly polarized light (Stokes $V$), which can still be observed even in heavily broadened lines. The high resolution of our observations allows us to resolve individual helium and metallic line profiles and unambiguously detect the presence of magnetic signatures in the observed spectra, depending on the relative noise level and amplitude of the signatures. Furthermore, for most magnetic configurations, we are able to detect a Zeeman signature even when the net line-of-sight component of the magnetic field (the longitudinal field $B_\ell$) is consistent with zero. This is not possible with low-resolution instruments that cannot typically resolve individual line profiles, and therefore only rely on the $B_\ell$ measurement for detection purposes.

\begin{figure}
\centering
\includegraphics[width=5.25in]{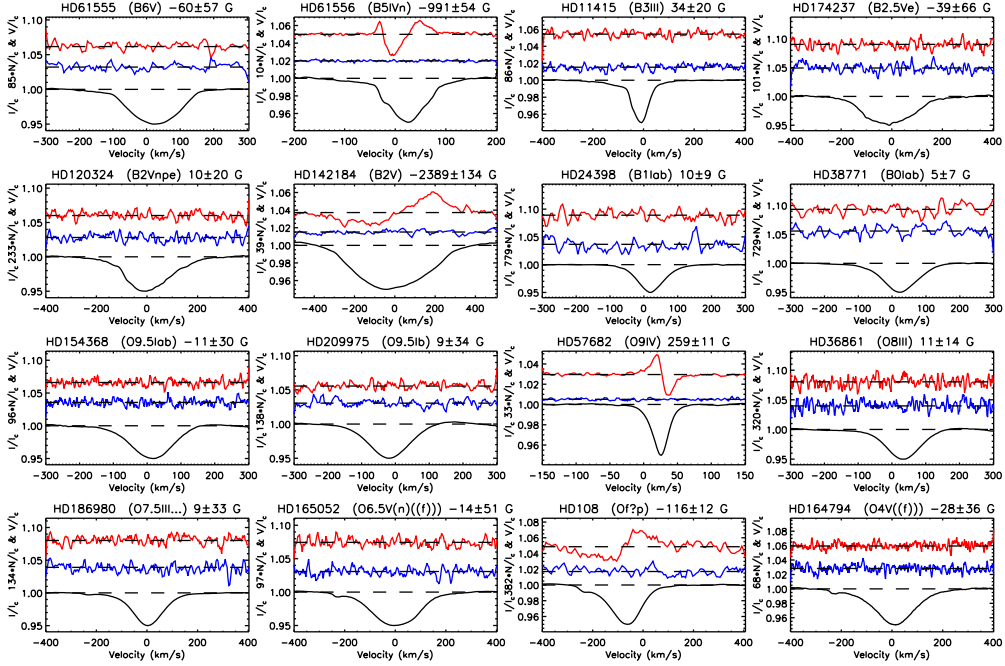}
\caption{Mean LSD circularly polarized Stokes $V$ (top curve, red), diagnostic null (middle curve, blue) and unpolarized Stokes $I$ (bottom curve, black) profiles for a small sample of the Survey Component observations. The spectral lines have all been scaled such that the unpolarized absorption lines have a depth that is 5\% of the continuum and the null and Stokes $V$ profiles have been shifted and amplified by the indicated factor for display purposes. A magnetic field is detected if there is excess polarization signal detected within the line profile of the Stokes $V$ spectrum. Clear magnetic Zeeman signatures are visible in HD\,61556, HD\,142184, HD\,57682, and HD\,108 while no Zeeman signature is present in the other stars. Also included are the spectral type and computed longitudinal field for each LSD profile.}
\label{lsd_examp}
\end{figure}

Since the detected magnetic fields in most hot, massive stars are relatively weak, we require S/Ns on the order of $\sim$10\,000 to detect Zeeman signatures in individual optical spectral lines of typical O- or B-type stars. In order to increase the S/N and enhance our sensitivity to weak Zeeman signatures, we employ the Least-Squares Deconvolution (LSD) multi-line technique of \citet{don97} to produce a single mean profile from the unpolarized light, circularly polarized light and also from the diagnostic spectrum (this null spectrum diagnoses the presence of spurious signatures resulting from instrumental or other systematic effects). The S/N increase is dependent on the number of spectral lines used in the analysis. For comparison with other studies, we also compute the first order moment of Stokes $V$, which was shown to be proportional to $B_\ell$ \citep{rees79}. Example LSD profiles are shown in Fig.~\ref{lsd_examp} for several different SC stars with different spectral types.

\section{Preliminary Results}
As of July 11th, 2011, the MiMeS SC has observed approximately 166 stars with ESPaDOnS and Narval. This corresponds to 2176 exposures (for certain targets we combine multiple sequences of observations to improve the S/N without saturating the CCD), resulting in 544 polarized spectra (4 exposures are required for a single polarimetric observation). This consumed roughly 228 hours of telescope time. The data are of extremely high precision, with a mean peak S/N per 1.8\,km\,s$^{-1}$ pixel in the unpolarized spectra of $\sim$1300. After performing a preliminary LSD analysis of these spectra, we measure an average S/N of $\sim$14\,000 in the mean LSD Stokes $V$ profiles.

\begin{figure}
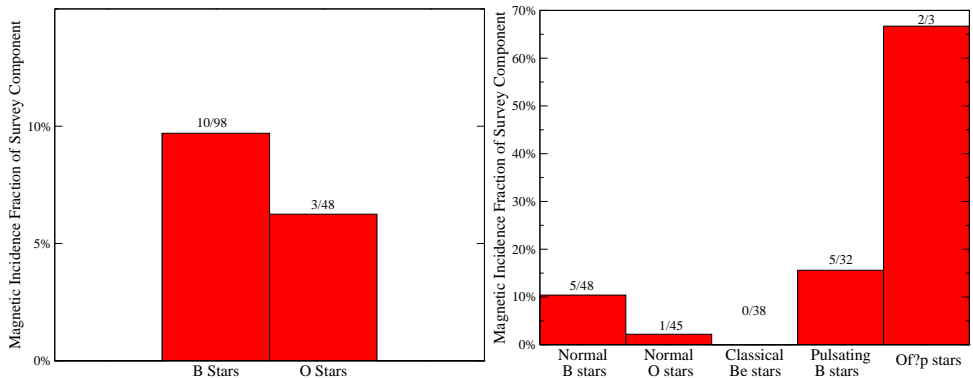

\centering
\includegraphics[width=2.5in]{cat_type_det_hist2.eps}
\includegraphics[width=2.5in]{cat_type_det_hist.eps}
\caption{Incidence fraction of magnetic stars relative to the total indicated sample of all stars observed as part of the MiMeS survey and related PI programs. Our statistics indicate a $\sim$8\% incidence rate of magnetism amongst all massive stars in our sample.}
\label{stat_res}
\end{figure}

From our preliminary statistics we find that approximately 8\% of all O- and B-type stars observed as part of the SC and related PI programs host detectable magnetic fields. If we look at the incidence rate of magnetic fields amongst just O- or B-type stars (left panel of Fig.~\ref{stat_res}) we find that $\sim$10\% of B-type stars host magnetic fields, while only 6\% of O-type stars are found to be magnetic. 

If we further classify the observed stars based on their spectroscopic or other fundamental properties (right panel of Fig.~\ref{stat_res}) we find that the peculiar class of massive, emission line Of?p stars stands out, with roughly 66\% (2/3) of these stars observed to be magnetic. These are small number statistics but if we include another Of?p star, which was previously found to be magnetic  \citep[HD\,191612,][]{don06}, and is being observed as part of the TC, we find that 3/4 of these stars that were observed with ESPaDOnS are found to be magnetic. Given this high detection rate, it may well be that a magnetic field was not detected in the other observed star since this star is too faint and the resulting observation has a significantly lower S/N compared to the other Of?p stars.

We also point out the class of pulsating B-type stars that consists of the Slowly Pulsating B-type (SPB) stars and the $\beta$ Cep pulsators. These stars are of particular interest since it has been claimed that upwards of 40\% of these stars have been found to be magnetic from low-resolution observations \citep[e.g.][]{hub06, hub09}. Our measurements are in conflict with these results as we only find detectable magnetic fields in about 16\% of the observed stars, which is similar to the incidence fraction ($\sim$10\%) that we obtain in normal B-type stars. For a few of these stars we have obtained multiple observations and have carried out a more detailed analysis \citep{shultz11}.

Lastly, another interesting preliminary result is the lack of detected fields in any classical Be star. We follow the definition of classical Be stars as B-type stars close to the main sequence that exhibit line emission over the photospheric spectrum, resulting from circumstellar gas that is confined to an equatorial Keplerian decretion disk \citep{port03}. This is in contrast to the common Be star classification, which is typically defined as a B-type star that has shown emission during at least one point in its life. This distinction allows us to investigate the potential role of magnetic fields in the formation of such circumstellar disks.

The lack of detections in classical Be stars does not appear to be a result of any systematics. Magnetic fields have been found in other rapidly rotating B-type stars observed in the SC and we have sampled a population of classical Be stars for which we should have detected a few magnetic stars if we assume a similar incidence fraction ($\sim$10\%) and magnetic field properties to the normal B-type stars. We also believe that the additional contribution of emission should not affect our results since we detect magnetic fields in other emission line O- and B-type stars (that do not fall into the classical Oe or Be star classification, e.g. Of?p stars), which indicates we are capable of observing Zeeman signatures in stars with emission.

\section{Summary}
In summary, the MiMeS collaboration has detected a total of 14 new magnetic massive stars as part of the MiMeS LP. Of these 14 stars, 3 are O-type stars, which doubles the number of magnetic O-type stars known prior to the start of the MiMeS Project. Our preliminary statistics suggest that $\sim$8\% of all hot, massive stars host strong, large-scale magnetic fields, which qualitatively agrees with the 5-10\% incidence rate that is found for intermediate mass stars \citep{power07} suggesting a similar mechanism for the magnetic field generation. We do not detect large-scale magnetic fields in any classical Be star, likely indicating that strong, large-scale fields do not take part in the generation of the equatorial decretion disks. Lastly, we tentatively suggest that we have identified a class of magnetic O-type stars, the Of?p stars, with fields being detected in all stars of this class observed with reasonable S/Ns.

\end{document}